\documentclass[conference,compsoc]{IEEEtran}

\usepackage{graphicx}
\usepackage{hyperref}
\usepackage{xcolor}
\usepackage{multirow}
\usepackage{amsmath}
\usepackage[numbers]{natbib}
\usepackage{xcolor,colortbl}

\usepackage{enumitem}
\setlist[itemize]{left=0.5em, itemsep=0.2em}

\newcommand{\etc}{\textit{etc. }}
\newcommand{\ie}{\emph{i.e., }}
\newcommand{\eg}{\emph{e.g.,}}
\newcommand{\technical}{\cellcolor[HTML]{33d4ff}x}
\newcommand{\social}{\cellcolor[HTML]{ffc878}x}
\newcommand{\environmental}{\cellcolor[HTML]{50cf5c}x}
\newcommand{\economic}{\cellcolor[HTML]{ff7878}x}
\title{Using Sustainability Impact Scores for Software Architecture Evaluation}

\makeatletter
\newcommand{\linebreakand}{%
  \end{@IEEEauthorhalign}
  \hfill\mbox{}\par
  \mbox{}\hfill\begin{@IEEEauthorhalign}
}
\makeatother

\author{\IEEEauthorblockN{Iffat Fatima, Patricia Lago}
\IEEEauthorblockA{Vrije Universiteit Amsterdam\\
Amsterdam, The Netherlands\\
\{i.fatima, p.lago\}@vu.nl}
\and 
\IEEEauthorblockN{Vasilios Andrikopoulos}
\IEEEauthorblockA{University of Groningen \\
Groningen, The Netherlands\\
v.andrikopoulos@rug.nl}
\and
\IEEEauthorblockN{Bram van der Waaij}
\IEEEauthorblockA{ Advanced Computing Engineering TNO \\
Groningen, The Netherlands\\
bram.vanderwaaij@tno.nl}
}

\begin{document}

\maketitle

\begin{abstract}

For future regulatory compliance, organizations must assess and report on the state of sustainability in terms of its impacts over time. Sustainability, being a multidimensional concern, is complex to quantify. This complexity further increases with the interdependencies of the quality concerns across different sustainability dimensions. The research literature lacks a holistic way to evaluate sustainability at the software architecture level. With this study, our aim is to identify quality attribute (QA) trade-offs at the software architecture level and quantify the related sustainability impact. To this aim we present an improved version of the Sustainability Impact Score (SIS), building on our previous work. The SIS facilitates the identification and quantification of trade-offs in terms of their sustainability impact, leveraging a risk- and importance-based prioritization mechanism. To evaluate our approach, we apply it to an industrial case study involving a multi-model framework for integrated decision-making in the energy sector. Our study reveals that technical quality concerns have significant, often unrecognized impacts across sustainability dimensions. The SIS coupled with QA trade-offs can help practitioners make informed decisions that align with their sustainability goals. Early evaluations can help organizations mitigate sustainability risks by taking preventive actions.

\end{abstract}


\maketitle
%%%%%%%%%%%%%%%%%%%%%%%%%%%%%%%%%%%%%%%%%%%%%%%%%%%%%%%%%%%%%
%                    Introduction
%%%%%%%%%%%%%%%%%%%%%%%%%%%%%%%%%%%%%%%%%%%%%%%%%%%%%%%%%%%%%
\section{Introduction}

Sustainability has emerged as an important concern in the software industry, especially with the growing awareness of its economic, environmental, and social impacts~\cite{Venters2023}. With regulations like the Corporate Sustainability Reporting Directive (CSRD)~\cite{csrd}, organizations are now bound in many European countries to report on sustainability~\cite{Hummel-2024}. However, sustainability, that is a multidimensional concern, is complex to quantify~\cite{Hummel-2024}. Sustainability can be framed as a quality property of software~\cite{Lago_2015-ssq} represented by quality attributes (QAs) that have impacts across multiple dimensions~\cite{Lago-action}. Thus, sustainability can be represented in various dimensions, such as economic (Ec), environmental (En), social (S), and technical (T)~\cite{Lago_2015-ssq}, with direct, enabling and systemic impacts~\cite{Hilty_2015}. As the number of QAs increases, it becomes more complex to evaluate sustainability.

In software engineering, early-stage decisions play a vital role in determining the long-term quality of the software~\cite{sheard_2019}. These decisions are made primarily at the level of software architecture (SA), which \citet{Jansen2005} conceptualize as \textit{a set of design decisions} that shape the behavior and structure of the software throughout its lifecycle. The evaluation of SA helps examine the conformity of SA with expected quality attributes (QA) and analyzes the potential trade-offs resulting from architectural decisions \cite{bass2012software}. It enables early identification of risks and helps to make informed decisions that maintain the quality of the system over time. This proactive approach can help in anticipating challenges and refining architectural choices to optimize the software's overall sustainability. 

In our previous work~\cite{2023_Fatima_SLR}, we presented an overview of software architecture (SA) evaluation methods to identify support for sustainability assessment. Our results showed that the Architecture Trade-off Analysis Method (ATAM) and its variants are the most prevalent in the literature. However, a lack of sustainability-focused SA evaluation approaches is observed. Based on previous work~\cite{2023_Fatima_SLR}, we designed and evaluated an SA Assessment Method for sustainability \cite{2024_Fatima_SA_Assessment_Method}. We introduced a Sustainability Impact Score (SIS) to quantify the sustainability impact of design decisions. However, the current design of the SIS poses certain limitations elaborated in Section \ref{subsec:limitations}. 

In this study, we provide a new version of SIS based on risk- and importance-based prioritizations. We use this SIS to evaluate the SA in an industrial case study. Our subject case is the architecture of the software system that provides decision-making support to the energy sector in the Netherlands. There are a large number of stakeholders in this project, ranging from municipalities, energy companies, service providers, and the general public. The project uses a large number of models as black boxes that come from different providers. One of the quality concerns that the project aims to address is the interoperability of these models using workflows. In this study, we explore how fundamentally technical concerns of the project impact the environmental, economic, and social dimensions of sustainability. Lastly, we discuss the benefit that our approach brings to the practitioners in the industry for creating sustainable digital solutions that will help comply with future sustainability regulations. 
%%%%%%%%%%%%%%%%%%%%%%%%%%%%%%%%%%%%%%%%%%%%%%%%%%%%%%%%%%%%%
%                   Proposed Approach
%%%%%%%%%%%%%%%%%%%%%%%%%%%%%%%%%%%%%%%%%%%%%%%%%%%%%%%%%%%%%
\section{The Journey So Far}
\label{sec:journeysofar}
In this section, we outline the techniques underlying the SIS and their potential limitations.   
%%%
\subsection{Dependency Matrix}
\begin{figure}[!htbp]
    \centering
    \includegraphics[width=0.6 \linewidth]{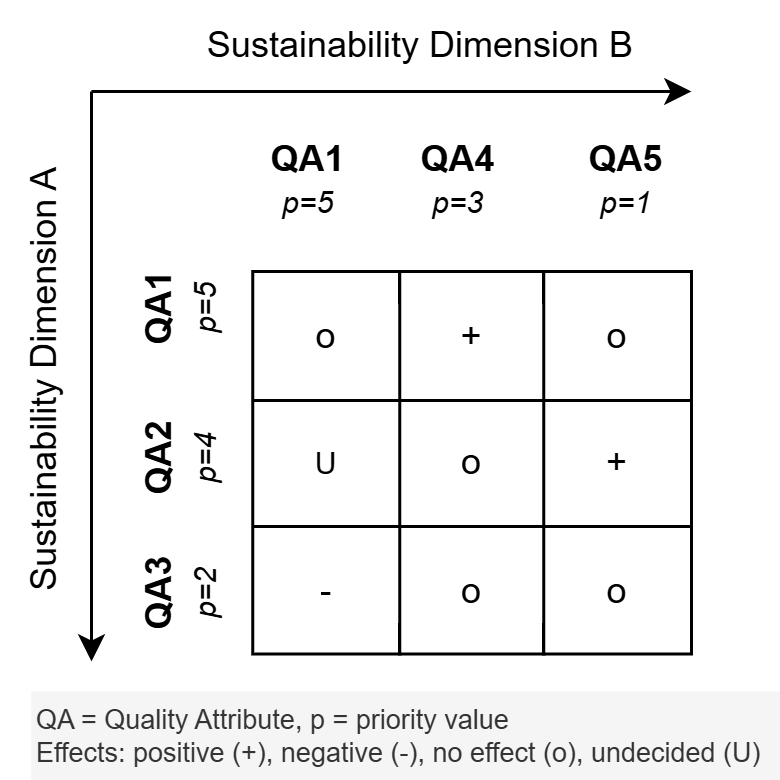}
    \caption{DMatrix Overview}%
    \label{fig:dmatrix-overview}%
    \vspace{-1em}
\end{figure}

We used the Dependency Matrix (DMatrix) from \citet{CondoriFernandez2024-dmatrix} to identify the dependencies between QAs in different sustainability dimensions. Fig.~\ref{fig:dmatrix-overview} shows an overview of the DMatrix structure.
Here, we elaborate on the different elements of the DMatrix and the source of information for each step. 

\textbf{Sustainability Dimensions.}
The DMatrix is a representation of a pair of sustainability dimensions (\eg~technical-economic, technical-environmental, technical-social, and so on) and their corresponding QAs. Each axis corresponds to a sustainability dimension. As a result, a single DMatrix encapsulates the effects of QAs in one sustainability dimension on another. The effects on QAs depend on the specific definition of QA within each sustainability dimension. 

\textbf{Quality Attributes and Effects.}
The DMatrix represents the effects of QAs represented along the vertical axis on those depicted on the horizontal axis, as positive, negative, neutral/undecided. 
%%%%%%%%%%%%%%%%%%%%%%%%%%%%%%%%%%%%%%%%%%%
%%%%%%%%%%%%%%%%%%%%%%%%%%%%%%%%%%%%%%%%%%%
%%
\subsection{Sustainability Impact Score}
\label{subsec:sis}
%%%%%%%%%%%%%%%%%%%%%%%%%%%%%%%%%%%%%%%%%%%
\begin{figure*}[hbtp]
    \centering
    \includegraphics[width=0.85\textwidth]{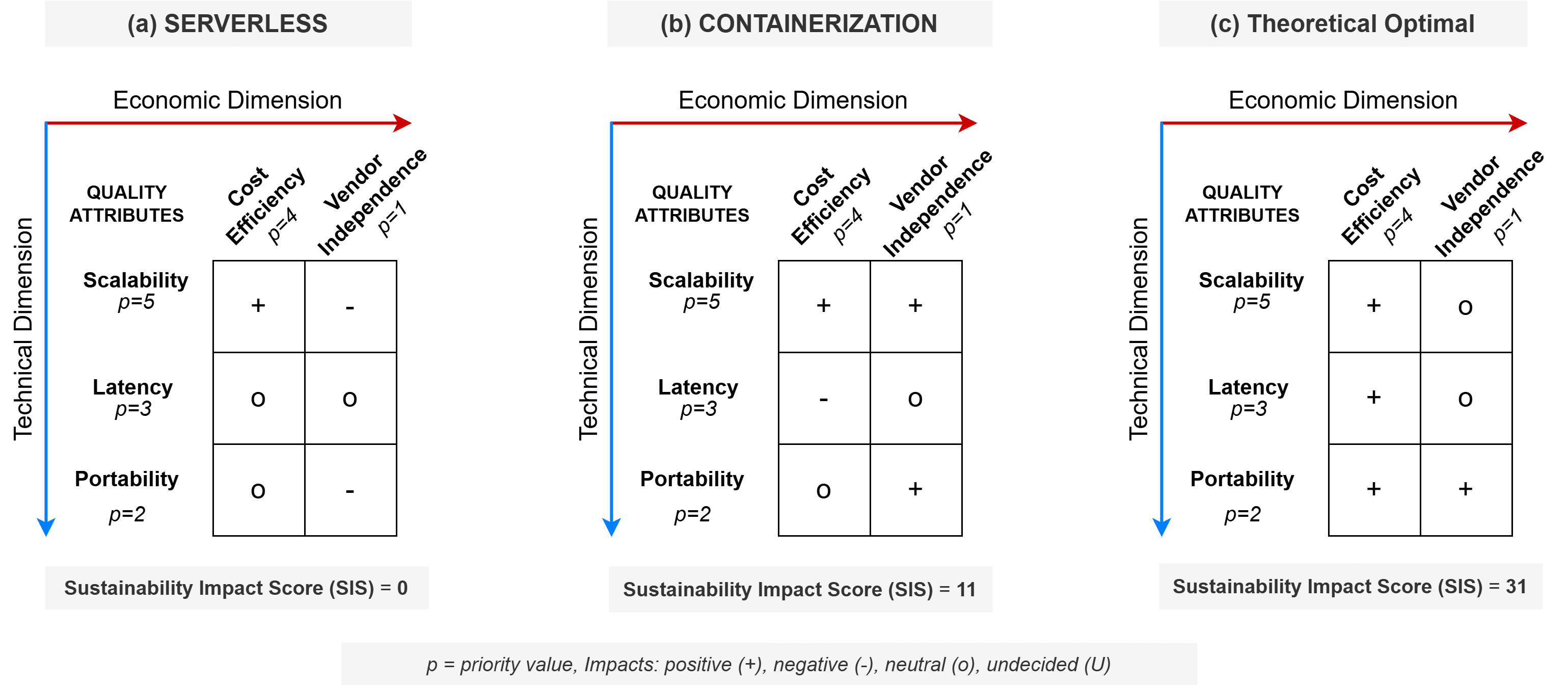}
    \caption{DMatrices for a. Serverless (left) b. Containerization (middle) c. Theoretical Optimal (right)}
        \label{fig:examples}
\end{figure*}
%%%%%%%%%%%%%%%%%%%%%%%%%%%%%%%%%%%%%%%%%%%
In our previous work~\cite{2024_Fatima_SA_Assessment_Method}, we introduced the concept of a SIS, which quantifies the impact of QAs within the technical dimension \texttt{T} about QAs within other sustainability dimensions \texttt{dim(Ec, En, S)}. 
Taking inspiration from the weighted sum model~\cite{wiley_2021-gf}, this score uses QA priority values as weights to assess the impact of QA. Subsequently, it computes the cumulative sum of the products of priority and the corresponding impact values, as depicted in Eq.~\ref{eq:sis-og}. 
%%%%%%%%%%%%%%%%%%%%%%%%%%%%%%%%%%%%%%%%%%%
\begin{equation}
\text{SIS}_{T, dim} = \sum_{\substack{i=1 \\ j=1}}^{n,m} (\text{Priority}_{T\_i} \text{ + } \text{Priority}_{dim\_j}) \times \text{Impact}_{ij}
\label{eq:sis-og}%
\end{equation}
 \text{\footnotesize where n=total QAs per dimension T, m=total QAs per dimension {dim},}\\ 
 \text{\footnotesize{dim} $\in$ \{Ec, En, S\}, impact = \{+1, -1, 0\}}   
%%%%%%%%%%%%%%%%%%%%%%%%%%%%%%%%%%%%%%%%%%%

\textbf{Example.}
Here, we consider the example of choosing between serverless vs containerization for data pipelines from the work of \citet{lekkala2023containerization}. We pick the following set of evaluation criteria for this case: scalability, cost-efficiency, latency, portability, and vendor independence with priorities of 5, 4, 3, 2, and 1, respectively (from highest to lowest). We map these QAs to different sustainability dimensions based on their definitions. We place scalability, latency, and portability under the technical dimension, while we place cost-efficiency and vendor independence under the economic dimension. We create DMatrices for both options and calculate the SIS based on Eq. \ref{eq:sis-og}. 

\textbf{Inter-QA trade-offs.}
Fig.~\ref{fig:examples} (a) shows that for serverless architecture, scalability has a positive effect on cost efficiency due to auto-scaling capabilities. However, this positive impact comes at the cost of losing vendor independence, which can have negative economic effects. Portability has a negative effect on vendor independence, as serverless systems induce vendor lock-in, making portability hard. 

Fig.~\ref{fig:examples} (b) shows that for containerization-based architecture, scalability support has a positive effect on cost efficiency while still maintaining vendor independence, as compared to serverless architecture. However, the latency might be higher for this approach, creating a negative effect on cost efficiency. Containerization supports portability due to vendor independence, hence leading to a positive effect.  

\textbf{Intra-dimension trade-off.}
On top of the QA tradeoffs, SIS indicates the support for the sustainability dimension pairs. In this case, Fig. \ref{fig:examples} shows that containerization (SIS=11) has a higher SIS compared to serverless (SIS=0). Given the relatively higher SIS value, choosing containerization is the more sustainable design option considering the technical and economic dimensions. Similarly, we can create DMatrices for different dimension pairs to observe related trade-offs and sustainability impact. 
\subsection{Limitations of SIS}
\label{subsec:limitations}
With our experience in performing SA evaluations in our previous work~\cite{2024_Fatima_SA_Assessment_Method}, SIS presented with limitations like (i) SIS was based on ranking-based prioritization, which is prone to bias and subjectivity; (ii) SIS was only comparable within same dimension pair (\eg T-Ec with E-Ec) and not comparable across different dimension pairs (\eg T-Ec with T-En), and (iii) SIS had a different upper and lower limit for different dimension pairs, making the numbers incomparable. 
%%%%%%%%%%%%%%%%%%%%%%%%%%%%%%%%%%%%%%%%%%%%%
%                    Evaluation
%%%%%%%%%%%%%%%%%%%%%%%%%%%%%%%%%%%%%%%%%%%%%
\section{Proposed Improvements in SIS}
\label{sec:improvements}%
Based on the identified limitations, in this work, we proposed a new definition of SIS with a risk and importance-based prioritization scheme. We used these improvements in an industrial case, as elaborated in Section \ref{sec:methodology}.

\textbf{SIS for all dimension pairs.}
We modified SIS by \citet{2024_Fatima_SA_Assessment_Method} to calculate SIS across any pair of dimensions and redefined its representation in Eq.~\ref{eq:sis-new}
%%%%%%%%%%%%%%%%%%%%%%%%%%%%%%%%%%%%%%%%%%%%%
\begin{equation}
\text{SIS}_{dim1, dim2} = \sum_{\substack{i=1 \\ j=1}}^{n,m} (\text{Priority}_{dim1\_i} \text{+} \text{Priority}_{dim2\_j}) \times \text{Impact}_{ij}
\label{eq:sis-new}%
\end{equation}
\text{\scriptsize where n=total QAs per dimension dim1, m=total QAs per dimension dim2}   \\
\text{\scriptsize \texttt{dim} $\in$ \{Ec, En, S, T\}, impact = \{+1, -1, 0\}}\\   
%%%%%%%%%%%%%%%%%%%%%%%%%%%%%%%%%%%%%%%%%%%%%
\textbf{Risk and Importance based Prioritization.}
Risk is an important factor when it comes to SA evaluation in the industry. Hence, taking inspiration from ATAM~\cite{S1}, we employed a risk-averse technique for prioritization of QAs. 
In ATAM, utility trees offer a structured approach to effectively map the business objectives of a system into specific QA scenarios. These scenarios are prioritized based on two factors which are (i) the level of importance for the success of the system, and (ii) the level of risk to achieve quality~\cite{S1}.

\textit{Introducing the Utility Matrix.}
The utility tree limits the scenarios to a single QA. In the real world, QAs and scenarios do not have a one-to-one relationship, rather they have a many-to-many relationship. We represent this many-to-many relationship using a matrix with QAs on the left and scenario IDs at the top. We iteratively build this matrix. Using the utility tree approach we identify the scenarios important to each QA. In the second iteration, we link the scenarios to other QAs if they are inter-related. We call this a \textit{utility matrix}.
% s(template provided in \cite{rep-pkg}).
%%%%%%%%%%%%%%%%%%%%%%%%%%%%%%%%%%%%%%%%%%%%%

\textit{Prioritization using Utility Matrix.}
The priority values described in ATAM are qualitative. However, to calculate SIS, a single numerical value for priority is required. This need is addressed in Eq.~\ref{eq:weighted-priority-score}, which defines a priority function calculated as a weighted sum of importance and risk levels. The weights for these levels can be determined by the system's stakeholders while developing the utility matrix. We used max-min normalization to bring the priority values between a range of 0.1 and 1. These priority values will be used to calculate the SIS for the alternatives represented in each scenario. 
\begin{equation}
\text{Priority}_{S_i} = f(I, R) = w_I \cdot I + w_R \cdot R
\label{eq:weighted-priority-score} %
\end{equation}
\text{\scriptsize $ S_i=\textit{ith Scenario}, I=\textit{Importance}, R = \textit{Risk},\; \\ I,R \in \{\text{High=3}, \text{Medium=2}, \; $}
\text{\scriptsize $ \text{Low=1}\}, w_I=\textit{importance weight}, w_R = \textit{risk weight}, 0 \leq w_I, w_R \leq 1 \} \; $}\\
%%%%%%%%%%%%%%%%%%%%%%%%%%%%%%%%%%%%%%%%%%%%%
\textbf{Relative comparison of SIS.}
To objectively conclude the SIS values, we create DMatrices showing theoretical optimal cases for a relative comparison. Such a DMatrix represents the effects to obtain a maximum SIS value which is both theoretically possible and optimal from a utility perspective. A higher SIS value indicates support for the particular dimension. So, $SIS_{T-Ec}$ shows the support of the economic dimension based on the effects of QAs in the technical dimension.
% Figure~\ref{fig:examples} (c) displays a DMatrix that shows an optimal scenario where maximum positive results can be achieved.

In a theoretically optimal case, QAs have minimal or no negative effects on another. In this theoretical optimal case, SIS (in Fig.~\ref{fig:examples} (c)) can be used as a comparative benchmark to assess whether the SIS values in Fig.~\ref{fig:examples} (a and b) impact the economic dimension and, if yes, by what magnitude. We can calculate this as a relative percentage value where the theoretical optimal represents 100\%. All other SIS percentages reflect proximity to this benchmark. Negative percentages indicate underperformance relative to the ideal. For example, in the case of the example in Section \ref{sec:journeysofar}, we saw that serverless provided a 0\% sustainability support as compared to the theoretical optimal of 100\%. However, containerless provided sustainability support of up to 35.48\% as compared to the theoretical optimal, making it a better choice for reaping economic benefits. 
%%%%%%%%%%%%%%%%%%%%%%%%%%%%%%%%%%%%%%%%%%%%%%%%%%%%%%%%%%%%%
%                   Study Design
%%%%%%%%%%%%%%%%%%%%%%%%%%%%%%%%%%%%%%%%%%%%%%%%%%%%%%%%%%%%%
\section{Case Study Design}
\label{sec:methodology}%
In this section, we elaborate on the steps to execute the case study towards evaluating our improved SIS proposal. The study was carried out together with one of the stakeholders in the project, who also took the role of architect among others, in the project. The choice of study participant is made based on the availability among team members and the level of knowledge about the MMvIB IT architecture. 

\textbf{Case Description.}
To clarify our approach, we illustrate it with an example. The subject is the MMvIB IT architecture~\cite{multimodelling-docs}. MMvIB stands for ``Multi modeling voor Integrale Besluitvorming'' (`Multi model for Integral Decision making', in English). 
MMvIB is a system utilized for planning and designing energy systems, which inherently involve high-stakes decision-making amidst many uncertainties. These decisions are often interdependent, contributing to the complexity of the system. To address this, various models are employed as sources of information for data inference, simulation, optimization, \etc However, given the intricate nature of these decisions, no single model can independently provide comprehensive information. Consequently, it becomes necessary to infer multiple models in conjunction, leading to the concept of interworking models known as a multi model. Depending on the different insights required, several multi models might be needed.

\textbf{Exploring the Problem Space.}
Together with the study participant, we identified the challenges of the MMvIB platform and the architectural approaches. Then, we identified the quality concerns to overcome the identified challenge. Finally, we prioritized these concerns based on the importance and risk levels identified by the participant. We formulated a set of questions to explore this problem space together with our study participant. The questions are provided in our replication package~\cite{rep-pkg}. 

\textbf{Brainstorming using a Decision Map.}
We take the support of Decision Map (DMap) from the SAF Toolkit~\cite{Lago2024-SAF-Toolkit} to explore the problem space. The DMap helps to (i) identify architectural concerns in different dimensions of sustainability, (ii) identify the effects of these architectural concerns on each other (+1, -1, or 0), and (iii) identify the impact levels; direct, enabling, or systemic. This DMap helped us identify the missing QAs and the effects, later used in the DMatrices. 

\textbf{Mapping Quality concerns to sustainability dimensions.}
We took support from the Sustainability Quality Model of the SAF Toolkit~\cite{Lago2024-SAF-Toolkit} to define QAs according to sustainability dimensions.

\textbf{QA Prioritization.} The study participant chose the risk and importance levels for each QA as described in Section~\ref{sec:improvements}. To mitigate bias at the time of choosing these levels, we do not inform the participant that these values translate into the priority of QA. 

\textbf{Trade-off analysis using DMatrix.}
We took the support of the Dependency Matrix (DMatrix) ~\cite{CondoriFernandez2024-dmatrix} to document the effects of QAs in one dimension on the other, as explained in Section \ref{sec:journeysofar}. In addition, we also defined a DMatrix for the theoretical optimal case to obtain the maximum positive impacts. This helped us calculate the relative impact that a dimension had on the other. These DMatrices were computed for each dimension pair where there was at least one effect value (positive or negative).

\textbf{Sustainability Impact Score.}
We used the DMatrices to calculate a `SIS' for each pair of dimensions using Eq.~\ref{eq:sis-new} and \ref{eq:weighted-priority-score}. This gave us an indication of the trade-offs between sustainability dimensions. To enable relative comparison, we calculated the `Normalized SIS' using min-max normalization and converted the scores to a percentage value between 0 and 100\% using Eq.~\ref{eq:sis-norm}. This makes the SIS comparable across different dimension pairs.
\begin{equation} \footnotesize
    \text{Normalized SIS (\%)} = \frac{SIS - \min(SIS)}{TO(SIS) - \min(SIS)} \times 100
    \label{eq:sis-norm}%s
    \vspace{-0.5em}
\end{equation}%
\text{\scriptsize \textit{TO = Theoretical Optimal}} %
%%%%%%%%%%%%%%%%%%%%%%%%%%%%%%%%%%%%%%%%%%%%%%%%%%%%%%%%%%%%%
%                   Results
%%%%%%%%%%%%%%%%%%%%%%%%%%%%%%%%%%%%%%%%%%%%%%%%%%%%%%%%%%%%%
\section{Results}
\label{sec:results}%
This section outlines the study results based on collected data for evaluation and trade-offs.
%%%%%%%%%%%%%%%%%%%%%%%%%%%%%%%%%
\begin{figure*}[!ht]
    \centering
    \includegraphics[width=0.85\textwidth]{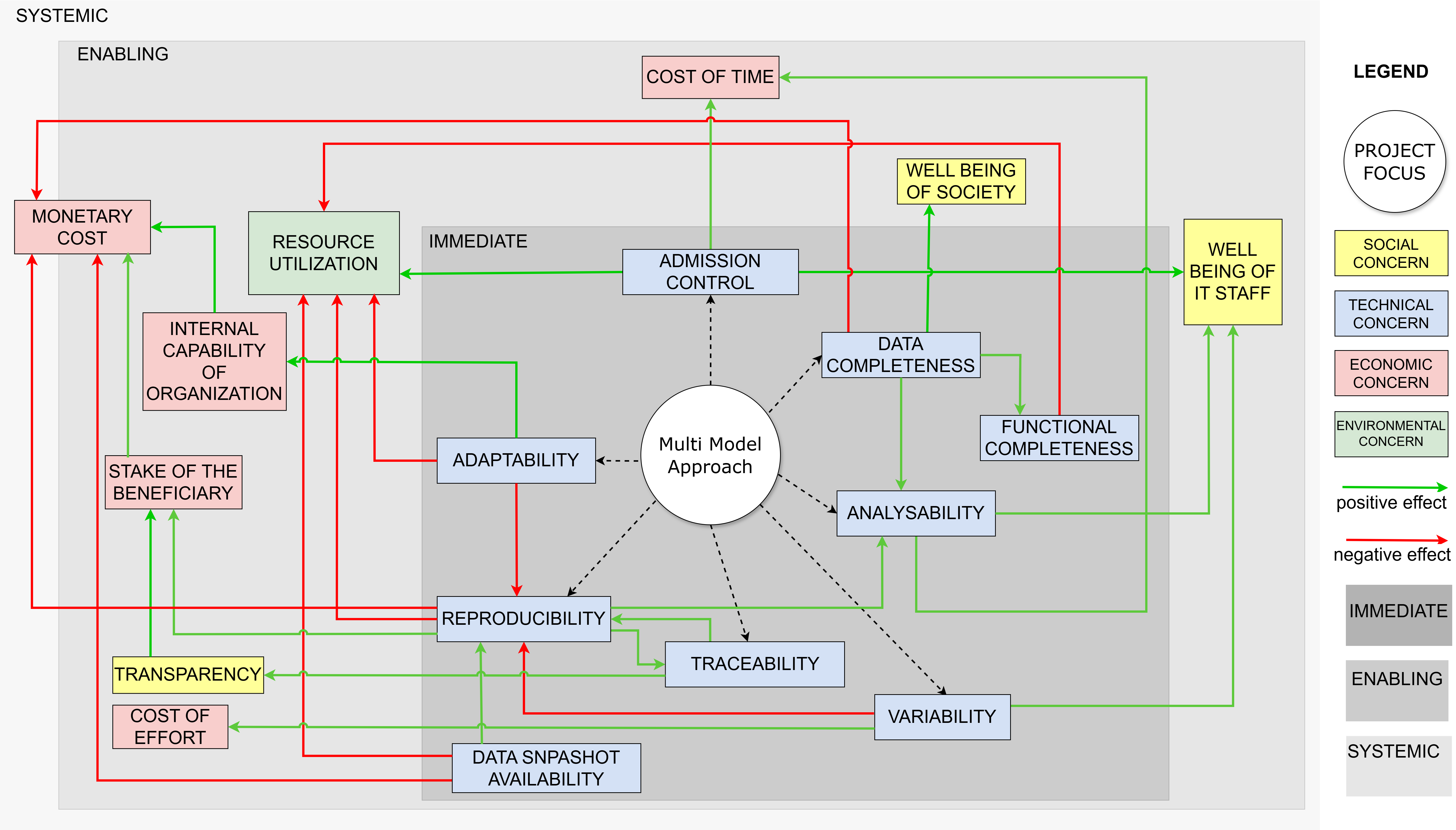}
    \caption{Decision Map showing quality concerns of the MMvIB IT Architecture using Multi Model Approach}
    \label{fig:dmap}%
\end{figure*}
%%%%%%%%%%%%%%%%%%%%%%%%%%%%%%%%%

\subsection{Exploring the Problem Space}
We started with the step \textit{Generation of data} from the SA Evaluation method by \citet{2024_Fatima_SA_Assessment_Method}. 
In our case study, interoperability was identified as a prominent challenge with the previous implementation of the system that MMvIB aims to support.
We identified the previous and current approaches for the MMvIB system (see Table~\ref{tab:arch-approaches}). The multi model approach was designed to meet interoperability needs. In this study, we compare these two approaches about a theoretical optimal case. 
%%%%%%%%%%% Architectural Approaches %%%%%%%
\begin{table}[!ht]
    \centering
    \caption{Architectural Approaches}
    \begin{tabular}{@{\hspace{0.5em}}p{1.6cm}@{\hspace{0.5em}}p{6.5cm}}
    \hline
        \textbf{Name} & \textbf{Description} \\ \hline
        Single Model & Manually run model(s) without automated workflows   \\ \hline
        Multi Model & Automate the process of running model(s) by creating a workflow which has intermediary models \\ \hline
    \end{tabular}
    \label{tab:arch-approaches}%
\end{table}

\textbf{Decision Map.} Based on interoperability needs, we used the requirement document to identify the quality concerns requirements and then used the DMap as a brainstorming tool to elicit other related concerns. The participant defined all quality concerns as QAs in the context of the project (see Table~\ref{tab:qas} for definitions). Together with our study participant, we created three DMaps for (i) the single model approach, (ii) the multi model approach, and (iii) the theoretical optimal case. Fig.~\ref{fig:dmap} illustrates the DMap for the multi model approach. It shows the quality concerns of the MMvIB IT Architecture. Other DMaps are provided in our replication package~\cite{rep-pkg}. We built the DMap by identifying the primary concerns (all technical), represented as blue boxes in Fig.~\ref{fig:dmap}. We then identified the other concerns that emerge as a consequence of supporting the primary concerns (social, economic, or environmental). The DMap shows that all of these concerns emerge as having enabling or systemic impacts. We then identify the type of effect on the concerns by iterating through all concerns one by one. These effects are identified as positive or negative. We used the same identified effects to build the DMatrices.

\textbf{Sustainability Quality Model.}
We use the quality concerns identified in the DMap and map these to the sustainability dimensions using the Sustainability Quality Model~\cite{Lago2024-SAF-Toolkit} (see Table~\ref{tab:qas})

\textbf{Prioritization.}
Table \ref{tab:qas} shows the list of QAs, the risk and importance levels identified together with the project stakeholder. As this is a single scenario case, we did not create utility matrices. Rather, we assigned equal weights (w1=w2=0.5) to both importance and risk levels for all QAs. Using Eq. \ref{eq:weighted-priority-score}, we calculated the priority value for each QA and normalized these values using min-max normalization. 

%%%%%%%%%%% QA Definitions %%%%%%%%%%%%%%%%%
\begin{table*}[!htbp]
    \centering
    \caption{QAs and their Definitions, Prioritization and Sustainability Quality Model}
    \begin{tabular}{|@{\hspace{0.5em}}p{2.5cm}|@{\hspace{0.5em}}p{9.3cm}|m{0.1cm}|m{0.1cm}|m{0.4cm}|m{0.4cm}|m{0.18cm}|m{0.18cm}|m{0.18cm}|m{0.18cm}|}
     \multicolumn{10}{l}{\textit{I=Importance, R=Risk, P=Priority, NP= Normalized Priority, Sustainability Dimensions: Ec=Economic, En=Environmental, S=Social, T=Technical}} \\ \hline
      \textbf{Quality Attribute} & \textbf{Definition} & \textbf{I} & \textbf{R} & \textbf{P} & \textbf{NP} & \textbf{Ec} & \textbf{En} & \textbf{S} & \textbf{T} \\ \hline
        Traceability & It refers to ensuring the entire workflow of the multi model approach is fully traceable, enabling the reproducibility of results. & 3 & 3 & 3.00 & 1.00 & ~ & ~ & ~ &  \technical  \\ \hline
        Adaptability & It refers to the capability for individuals or organizations to host their own instances of the workflow & 3 & 3 & 3.00 & 1.00 & ~ & ~ & ~ & \technical \\ \hline
        Variability & It refers to configuring the parameters for workflows and models' input/output to accommodate support for new parameters and changes. & 3 & 3 & 3.00 & 1.00 & ~ & ~ & ~ & \technical \\ \hline
        Admission Control & It refers to the ability to autonomously wait for a busy model to become available, distinguishing between busy and unavailable states. & 3 & 2 & 2.50 & 0.78 & ~ & ~ & ~ & \technical \\ \hline
        Data Completeness & It refers to the availability of all necessary and usable data for each model. & 3 & 2 & 2.50 & 0.78 & ~ & ~ & ~ & \technical \\ \hline
        Analysability & It refers to the ability to debug models to understand their semantics and functions for output generation, ensuring they are used as intended. & 2 & 1 & 1.50 & 0.33 & ~ & ~ & ~ & \technical \\ \hline
        Reproducibility & It refers to replicating the outputs of the workflow for validation purposes, ensuring all configurations, models, and datasets are versioned. & 2 & 1 & 1.50 & 0.33 & ~ & ~ & ~ & \technical \\ \hline
        Functional Completeness & It refers to the provision of outputs in line with the intended user's objectives. & 1 & 1 & 1.00 & 0.10 & ~ & ~ & ~ & \technical \\ \hline
        Data Snapshot Availability & It refers to the availability of data snapshot(s) used in a particular workflow in the past for reproducibility needs. & 1 & 1 & 1.00 & 0.10 & ~ & ~ & ~ & \technical \\ \hline
        Cost of Time & It refers to the time as a resource used to ensure the functionality of the IT system & 1 & 1 & 1.00 & 0.10 & \economic & ~ & ~ & ~ \\ \hline
        Cost of Effort & It refers to the effort as use of human resources to ensure the functionality of the system & 1 & 1 & 1.00 & 0.10 & \economic & ~ & ~ & ~ \\ \hline
        Monetary Cost & It refers to the gain or loss of capital as a result of the functioning IT system & 1 & 1 & 1.00 & 0.10 & \economic & ~ & ~ & ~ \\ \hline
        Stake of beneficiary & It refers to the stake of the beneficiaries of the system who can either be direct consumers or users & 1 & 1 & 1.00 & 0.10 & \economic & ~ & ~ & ~ \\ \hline
        Resource Utilization & It refers to the use of IT resources for the operation of the IT system & 1 & 1 & 1.00 & 0.10 & \economic & ~ & ~ & ~ \\ \hline
        Well being of IT staff & It refers to the time saved by the IT staff of the IT processes & 1 & 1 & 1.00 & 0.10 & ~ & \environmental & ~ & ~ \\ \hline
        Well being of society & It refers to the impact on the users as a community through the operation of this IT system & 1 & 1 & 1.00 & 0.10 & ~ & ~ & \social & ~ \\ \hline
        Internal Capability of Organization & It refers to improving the internal capability of the organization by developing their own workflows & 1 & 1 & 1.00 & 0.10 & ~ & ~ & \social & ~ \\ \hline
        Transparency & It refers to the degree to which the implementation of the workflow is visible to the beneficiaries  & 1 & 1 & 1.00 & 0.10 & ~ & ~ & \social & ~ \\ \hline
    \end{tabular}
    \label{tab:qas}%
    \vspace{-1em}
\end{table*}
\subsection{Synergies among Quality Attribute}
We also observed QA synergies, as some QAs positively (re)inforce other QAs. These cases are as follows. 
\begin{itemize}
    \item Traceability has a positive effect on Reproducibility (T), and Transparency (S). This Transparency (S) has a further positive effect on the Stake of the Beneficiary (Ec), which eventually leads to a positive effect on Monetary Costs (Ec). In this case, we observe that a primarily technical QA such as traceability has positive cascading effects across other sustainability dimensions. 
    \item Similarly, Traceability (T) also has a positive effect on Reproducibility (T), which in turn positively affects the Stake of the Beneficiary (Ec), creating a positive effect on Monetary Costs (Ec). 
    \item Ensuring conformance to Variability, Analysability, and Concurrency Control (T) has a positive effect on the well-being of IT staff (S) due to positive effects on Cost of Time (Ec) and Cost of Effort (Ec). 
\end{itemize}

\subsection{Trade-off Analysis}
We executed the step \textit{Evaluation of obtained data} from the SA Evaluation Method by \citet{2024_Fatima_SA_Assessment_Method} by using the data from the previous step. 
We created DMatrices from the collected data (QAs, priorities, and effects) to calculate the SIS. The DMatrices helped visualize the inter-QA trade-offs and the SIS helped visualize the inter-dimension sustainability trade-offs. 

\textbf{Inter-QA Trade-offs.} For inter-QA tradeoffs \textbf{within single dimension}, our results show that only the QAs in the technical dimension have trade-offs with other QAs in the technical dimension. Within the other three dimensions, \ie economic, environmental, and social, all QAs support each other or have no effect.

For \textit{single model approach}, only concurrency control, adaptability, data completeness, and functional completeness are supported. We did not observe any trade-offs between these QAs. For \textit{multi model approach,} we observed that reproducibility has maximum dependencies in terms of effects on and by other QAs. The trade-offs within the technical dimension are as follows.
\begin{itemize}
    
    \item {Adaptability} has a negative effect on Reproducibility. When organizations host instances with private models, reproducibility of the results becomes difficult. 
    \item {Variability} has a negative effect on Reproducibility. Allowing variations in the parameters of the model makes it difficult to reproduce the results as tracking the variation comes with an overhead. 
    
\end{itemize}

For inter-QA trade-offs \textbf{across dimensions}, our results show that resource utilization (En) is the most negatively affected QA as enabling impacts of the multi model system. The Well-being of IT staff (S) is the most positively affected QA. There are equal positive and negative effects on Monetary Costs as both enabling and systemic impacts. 

Although the single model approach supports a limited number of QAs, the trade-offs for the QAs are the same as in multi model approach. Therefore, we only discuss the trade-offs for the multi model approach in this section, as follows.  
\begin{itemize}
    \item Adaptability, Reproducibility, Data Snapshot Availability, and Functional Completeness (T) have negative enabling effects on Resource Utilization (En). The conformance to all these QAs requires additional resources. 
    \item Data completeness (T) has a negative systemic effect on Monetary Costs (Ec). To ensure functional completeness, data for all regions of the country is required. Most of these data come from third-party organizations that charge a large sum of money for data procurement or are proprietary and therefore are not openly accessible.    
    \item Data Snapshot Availability (T) has a negative enabling effect on Monetary Costs (Ec). Storing this data is essential for reproducibility. However, this also leads to a negative impact on Resource Utilization (En).
    \item Reproducibility (T) has a negative effect on Monetary Costs (Ec) which is strongly tied to the use of resources.
\end{itemize}

\textbf{Inter-dimension sustainability trade-offs.} We calculated SIS to identify the support of sustainability dimensions by the two architectural approaches and possible trade-offs. We translated the DMap into DMatrices for the following dimension pairs: T-T, T-Ec, T-En, T-S, and S-Ec. We did not consider other pairs because there were no effects identified for those pairs. We provide individual DMatrices with the SIS values in our replication package~\cite{rep-pkg}. 

Table~\ref{tab:sis} shows the SIS for both architecture approaches. Further, we also developed a DMap (followed by DMatrices) for a theoretical optimal case. Together with our study participant, we asked the question `Is it theocratically possible to reduce the negative impacts without creating a trade-off with high-priority technical QAs?' We answered this question by iterating each QA in the DMap and creating a DMap for a theoretical optimal case (provided in our replication package~\cite{rep-pkg}). We also calculated the SIS for this theoretical optimal case. This gave us an indication of the best-case scenario, which is theoretically possible. This allowed us to establish a comparison mechanism for the SIS values. 

\textbf{\textit{What do these SIS values mean?}}
Higher SIS values show a relatively higher support for a particular dimension. Negative values show that positive effects (if any) in a dimension are outweighed by negative effects. For now, we consider the effects of all QAs of the same magnitude (+1 or -1). However, one QA may affect another QA more positively or negatively than another QA. We leave this limitation to be addressed in our future work. 
The SIS shows that the multi model approach performs better across economic and social dimensions, while the single-model approach performs slightly less poorly across environmental dimensions. However, it should be noted that the environmental dimension is less affected by a single model as it does not support many high-priority QAs, such as reproducibility, adaptability, and data snapshot availability (which eventually affect the environmental dimension through increased resource utilization). 

To further make the comparison meaningful, we normalized the SIS values and compared them for both approaches with the theoretical optimal case (see Normalized SIS \% in Table~\ref{tab:sis}). The comparison was limited to the context of the architectural approaches and the available knowledge. The results show that the multi-model approach outperforms for economic dimension. Both approaches underperform across the T-En pair. Multi model approach has the maximum positive effect (100\%) across the T-S and S-Ec dimension pairs as compared to the theoretical optimal case. The normalization allowed us to compare the SIS values across dimensions which was not possible with the non-normalized values.
%%%%%%%%%%%%%%%%%%%%%%%%%%%%%%%%%%%%%%%%%%%%%%%%%%%%%%%%%%%%%
\begin{table}[!ht]
    \centering
    \caption{Relative Comparison of SIS}
    \begin{tabular}{p{2cm}m{1cm}m{1cm}m{1cm}m{1cm}}
    \hline
        \multirow{2}{*}{\textbf{\parbox{2cm}{Architectural Approach}}} & \textbf{$SIS_{T-Ec}$} & \textbf{$SIS_{T-En}$} & \textbf{$SIS_{T-S}$} & \textbf{$SIS_{S-Ec}$} \\ \cline{2-5}
        ~ & \multicolumn{4}{c}{\textbf{Non-normalized SIS}}  \\ \hline
         \text{Single Model} & -0.425 & -0.425 & 1.75 & 0 \\ \hline
          \text{Multi Model} & 2.425 & -1.05 & 4.375 & 0.2 \\ \hline
       \text{Theoretical Optimal}  & 3.3 & 1.15 & 4.375 & 0.2 \\ \hline
        \multirow{2}{2cm}{\textbf{~}} & \multicolumn{4}{c}{\textbf{Normalized SIS (\%)}} \\ \cline{2-5}
        \text{Single Model} & 0.00 & 28.41 & 0.00 & 0.00 \\ \hline
        \text{Multi Model} & 76.51 & 0.00 & 100.00 & 100.00 \\ \hline
        \text{Theoretical Optimal} & 100.00 & 100.00 & 100.00 & 100.00 \\ \hline
        \multicolumn{5}{l}{\footnotesize{\textit{Ec=Economic, En=Environmental, S=Social, T=Technical}}}
    \end{tabular}
    \label{tab:sis}%
    \vspace{-1em}
\end{table}
%%%%%%%%%%%%%%%%%%%%%%%%%%%%%%%%%%%%%%%%%%%%%%%%%%%%%%%%%%%%%
%                    Reflections
%%%%%%%%%%%%%%%%%%%%%%%%%%%%%%%%%%%%%%%%%%%%%%%%%%%%%%%%%%%%%
\section{Reflections}
\label{sec:reflections}
In this section, we discuss the results of our approach and how the SIS can provide support towards helping the software industry to meet their sustainability goals.

\textbf{Identification of the hidden enabling and systemic sustainability impacts.}
In the context of our case study, many inter-dimensional trade-offs arise from the enabling and systemic effects of implementing a QA, which, although not immediately visible, can have far-reaching consequences. For example, in our DMap (see Fig.~\ref{fig:dmap}) automating concurrency control mechanisms positively affects the social well-being of IT staff by reducing the need for prolonged waiting periods to rerun workflows. Additionally, it has a positive systemic impact, as the resource consumption of a waiting workflow is significantly lower than that of running multiple instances of models, which can increase the workload on cloud infrastructure. Although these impacts are indirect, DMaps facilitate the early identification of quality concerns and their effects on the SA evaluation process, facilitating more effective evaluation and management of potential trade-offs before they lead to undesirable outcomes. In the future, we aim to explore whether different effects should have a different magnitude based on the level of benefit or consequence, which would directly impact the SIS values. The study participant noticed that using DMaps in the SA evaluation process helped in visualizing the effects.

\textbf{ QA synergies and trade-offs.}
Our results showed synergies and trade-offs between QAs. This information is represented in the DMap and DMatrices as positive and negative effects. This information is important at the time of changing decisions, as we can observe the effect a change would have on other dimensions. For instance, looking at the DMap, if we reduce the reproducibility capacity, we would be able to reduce the negative effect of reproducibility on resource utilization. Hence, there will be an immediate negative effect on traceability and analysability, both of which are important for the success of the project, with the former being also high-risk. It will also negatively affect the stake of the beneficiary. We can quickly change these effects in the DMatrix to observe the changing SIS values and sustainability support. In this case, we choose to endure the tradeoff of reproducibility with resource utilization, as resolving it would lead to many other trade-offs (both immediate and enabling) with other high-priority QAs. It is important to note that this decision was made because resource utilization was not a quality requirement considered at the time of choosing these design decisions. It only appeared as a side effect during these assessment processes. Hence, it is important not only to identify the priority levels correctly but also to identify the relevant QAs that may affect the system early on in the project. Brainstorming approaches such as DMap can help mitigate this problem. The reliability of the SIS depends on the identification of all relevant quality concerns. In our experience, this step takes the longest time in the SA evaluation process.

\textbf{Inferring the SIS values.} Observing the non-normalized SIS values in Table \ref{tab:sis} for $SIS_{T-En}$ shows that the SIS can be negative for both approaches. This is because, naturally, the implementation of any technical system requires a certain amount of resources. To bring meaning to these values for relative comparison, we normalize them as a percentage between 0\% to 100\% using a theoretical optimal value. However, this can also skew the perspective if the negative impacts are extremely high. The theoretical optimal values help us compare the SIS values across sustainability dimensions. Since we do not have a lower bound for a SIS value, an open question remains: how bad are the lower values of SIS?
For an objective evaluation, organizations must evaluate their sustainability impacts based on certain benchmarks. For example, a company cannot produce more than \textit{x} grams of $CO_2$ emissions based on certain criteria. Without such benchmarks in place, we cannot compare the state of sustainability of digital solutions objectively.

\textbf{On learning.} Our reflective session with the participant revealed that the case study enhanced sustainability knowledge by addressing multiple dimensions. It also paved the way for knowledge sharing (\eg~\cite{Niemela2005}) and skill development~\cite{HELDAL2024}, enabling the participant to learn tools such as decision maps and SIS calculators.
The SIS offers a comprehensive overview of sustainability by assessing compliance with specific QAs and their impacts on other sustainability dimensions. Conducting the SA evaluation process early would serve as a proactive measure to ensure the sustainability of digital solutions. In addition, SIS can be used reactively to identify areas for improvement and address existing shortcomings. The participant in the case study noted that the SIS helped visualize the planned improvements from single to multimodel and interjected that such approaches can be more valuable if conducted upfront in the project. Benchmarks are needed for a well-rounded comparison of the state of sustainability of a digital solution. 
%%%%%%%%%%%%%%%%%%%%%%%%%%%%%%%%%%%%%%%%%%%%%%%%%%%%%%%%%%%%%
%                    Threats to Validity
%%%%%%%%%%%%%%%%%%%%%%%%%%%%%%%%%%%%%%%%%%%%%%%%%%%%%%%%%%%%%
\section{Threats to Validity}
\label{sec:ttv}
In this section, we present threats to the validity of this research using the categorization by \citet{wohlin2012experimentation} in the context of our study. 

\textbf{Construct Validity.}
The case study facilitator created a set of questions to systematize the data collection process. To mitigate risks of subjectivity and bias, we frame the questions without influencing the responses of the participant. 

\textbf{Internal Validity.}
The study faces potential biases from both the participant and the facilitator, which could affect internal validity. Personal biases or subjective judgments may influence the participant's interpretations of QAs and their effects. To mitigate these risks, we meticulously define the steps of the study's methodology to guide its execution. We adhere to the evaluation steps based on \citet{2024_Fatima_SA_Assessment_Method} and apply inquisitive brainstorming techniques for data collection and inference, ensuring that participant responses are not influenced. Iterative reviews of the results with the participant provide opportunities for improvement. Despite the limitations posed by a single stakeholder participant, we cross-checked the supporting documentation to ensure a complete understanding of the system. The SA is at a higher abstraction level, hiding complexity. We aim to evaluate at the design decision level to uncover hidden trade-offs.

\textbf{External Validity.}
Despite one case study, the SA evaluation method and the SIS quantification approach are architecture- and QA-agnostic. Hence, they can be applied to other SA evaluations. However, QAs, their effects, and the SIS values in this study cannot be generalized to other cases. Further, the involvement of multiple stakeholders will require a consensus-building step for the allocation of risk, importance, weight, and effect values.

\textbf{Conclusion Validity.}
The participant reflected on the rationale, decisions, and their implications to mitigate personal bias that could potentially affect reliability. To ensure conclusions were valid, the participant stayed within the project's scope avoiding going overboard with the prediction of sustainability impacts.

%%%%%%%%%%%%%%%%%%%%%%%%%%%%%%%%%%%%%%%%%%%%%%%%%%%%%%%%%%%%%
%                    Related Work
%%%%%%%%%%%%%%%%%%%%%%%%%%%%%%%%%%%%%%%%%%%%%%%%%%%%%%%%%%%%%
\section{Related Work}
\label{sec:relatedwork}
In this section, we provide an overview of the related work in the field of Software Architecture Assessment in terms of sustainability and multi-criteria decision-making. We finally elaborate on how our approach differs from the current approaches in terms of novelty and improvement. 

~\citet{S1} present the Architecture Trade-off Analysis Method (ATAM), which is the most prevalent scenario-based method in SA evaluation research with various light-weight variants~\cite{2023_Fatima_SLR}. 

On top of the need for inter-QA trade-off analysis, sustainability's multidimensional nature requires a decision-making mechanism that supports the analysis based on more than one evaluation criterion. Multi-Critical Decision Making (MCDM) is a widely used approach in software engineering decision-making. In the context of SA evaluation, many studies use the Analytic Hierarchy Process (AHP)~\cite{saaty1980analytic} for multi-criteria decision-making. AHP is used to quantify the impact of the evolvability of candidate architectural solutions on the subcharacteristics of evolvability~\cite{Breivold-2012-evolvabilityimpact} and the analysis of security risks~\cite{S39}. \citet{S35} present an AHP-based approach to analyze QAs in enterprise architecture (EA). We also observe Fuzzy AHP (FAHP) approaches toward EA analysis~\cite{S46}, security-specific analysis of SA by providing a security index~\cite{S59}, and quantitative maintainability analysis based on design principles~\cite{S61}. \citet{Cansado-2012} propose a weighted sum model (WSM) to evaluate EA scenarios. However, the study does not provide details about the use of WSM. We do not see any other studies in the software architecture evaluation literature using this approach for SA evaluation.

Although these MCDM methods are not presented with the intention of sustainability impact evaluation, their adoption to solve the problem of sustainability assessment is also not feasible. As explained in Section \ref{sec:journeysofar}, our sustainability evaluation requires support for a disjoint or overlapping set of evaluation criteria on each side of the matrix, which is normally the case in sustainability evaluation due to its multidimensional nature. Further, these methods are used for prioritization rather than inference of impact. In our approach we use a basic MCDM method, such as the weighted sum model (WSM)~\cite{wiley_2021-gf} in combination with the DMatrices. In our approach, we preserve sustainability assessment needs without adding the complexity of multiple-layered mathematical calculations. We aim to provide a simple and effective way to establish a SIS. We also provide a reusable template in our replication package \cite{rep-pkg}, which automates the quantification process and can be easily adopted by practitioners for their assessments. 

%%%%%%%%%%%%%%%%%%%%%%%%%%%%%%%%%%%%%%%%%%%%%%%%%%%%%%%%%%%%%
%             Conclusion and Future Work
%%%%%%%%%%%%%%%%%%%%%%%%%%%%%%%%%%%%%%%%%%%%%%%%%%%%%%%%%%%%%
\section{Conclusion and Future Work}
\label{sec:conclusion}
In this study, we perform an SA evaluation by comparing two architectural approaches and their relative comparison with a theoretical optimal case. We prioritize the evaluation criteria (or QAs) using a risk- and importance-based scheme. Our results include (i) identification of inter-QA trade-offs within a single dimension (T-T), (ii) identification of inter-QA trade-offs across different dimensions (T-Ec, T-S, T-En, and S-Ec), (iii) a sustainability impact score (SIS) to gauge support of a sustainability dimension, and (iv) a mechanism for relative comparison of SIS values across sustainability dimensions. The SIS can help practitioners in SA evaluation to preemptively mitigate sustainability risks in their projects. It can aid organizations in ensuring compliance with future regulations like CSRD. In the future, we aim to evaluate the SIS using multiple scenarios and a utility matrix, involving diverse stakeholders for an objective SA assessment.

%%%%%%%%%%%%%%%%%%%%%%%%%%%%%%%%%%%%%%%%%%%%%%%%%%%%%%%%%%%%%
%                    Acknowledgments
%%%%%%%%%%%%%%%%%%%%%%%%%%%%%%%%%%%%%%%%%%%%%%%%%%%%%%%%%%%%%
\section*{Acknowledgments}
This publication is part of the project SustainableCloud (OCENW.M20.243) of the research programme Open Competition by the Dutch Research Council (NWO).

\bibliographystyle{plainnat}
\bibliography{main}
\end{document}